# THE POSSIBILITIES OF THE QUANTUM MEMORY REALIZATION FOR SHORT PULSES OF LIGHT IN THE PHOTON ECHO TECHNIQUE

S.A.Moiseev, M.I.Noskov

Zavoisky Physical-Technical Institute of the Russian Academy of Sciences,
Sibirsky Trakt 10/7, Kazan, 420029, Russia,
E-mail: samoi@yandex.ru

The possibilities of recording, storage and reconstruction of short single photon wave packets in the photon echo technique are analyzed. The influence of the photon field and medium parameters on the quality and precision of the photon quantum state reconstruction is theoretically studied.

# THE POSSIBILITIES OF THE QUANTUM MEMORY REALIZATION FOR SHORT PULSES OF LIGHT IN THE PHOTON ECHO TECHNIQUE

S.A.Moiseev, M.I.Noskov

Zavoisky Physical-Technical Institute of the Russian Academy of Sciences,
Sibirsky Trakt 10/7, Kazan, 420029, Russia,
E-mail: samoi@yandex.ru

## 1.Introduction

Various quantum states of light and their interactions with atoms and molecules are interesting now for the investigation both of the fundamental problems of quantum mechanics, quantum optics and their applications, in particular, for quantum information. *Quantum memory (QM-)* is one of problems which is actual practically for all basic processes of quantum information. Now several proposals are investigated for solution of the QM-problem. Single atoms in cavity QED are the most simple and elegant quantum object which can be used for the quantum memory of single photon states [1, 2], but at the practical applications this proposal has a number of limitations for the quantum states of light. Particularly it is difficult to use the technique for a storage of the spectral broadened light and states with arbitrary number of photons. In this respect, using the coherent atomic ensembles could be especially preferable. Recently it was proposed a well known scheme of Raman-type interaction [3] for quantum storage. This scheme can work for the case of free space, nevertheless it does not allow to reversible store the quantum state and its retrieval reconstruction on the level of single photons. Most promising solution of the QM problem is proposed on the basis of electromagnetically induced transparency (EIT) effect [4]. Proof-of-principle experiments confirms the theoretical prediction [5,6] for the intensive laser pulses. Now a more general temporal properties and general possibilities of the EIT-technique [7, 8] as their present and nearest future perspectives [9, 10] are investigated theoretically. We note that the complete reconstruction of the arbitrary quantum state of light is especially difficult problems for experimental realisations of quantum memory processes. Experimental implementation of the QM process basing the coherent atomic ensembles on the level of the weak quantum field is now a challenging problem, where the separation of the weak quantum fields from the intensive laser fields which excite additionally the atoms in the EIT-technique is especially difficult and is needed in new ideas. It is important to develop the realisation of such QM-processes for short pulses of quantum fields optical field that is difficult to perform using an usual EIT-technique elaborated for the narrow spectral fields.

Recently, we have proposed the QM technique using the novel variant of the photon echo [11] , based on the reversibility properties inherent in an atomic gas of three-level resonant atoms with Λ-configuration of atomic transitions. This technique has some new advantageous properties as compared to the EIT-effect, which can be useful for the solution of the QM problem. In this paper we theoretically study the possibilities of our technique for realization of quantum memory for short single photon wave packets. This emphasis is determined by the necessity to suppress the negative role of relaxation processes in the medium.

In conclusion we discuss the advantages of the proposed technique and its potential for realization of quantum memory.

## 2. Basic model

Following [11] the single photon wave packet enters from the left (see Fig.1) the gas tube of N three-level atoms localised between the coordinates Z=0 and Z=0 and is absorbed in it on the transition $|1\rangle$ - $|3\rangle$. The first laser pulse transfers the single photon excitations from the level $|3\rangle$ to the long lived level $|2\rangle$ at t = $t_1$. After the storage the second pulse returns the saved excitation onto the level $|3\rangle$ at t = $t_2$. Following the time delay $t_1$, after the second laser pulse the atomic dipoles will rephased and the medium will irradiate the photon in the echo signal. It is important that the probability of the photon emission can be close to unit since this process is adjusted to be reversible to the absorption process of the initial photon. Below we consider stage by stage interaction of the stored single photon wave packet with the medium, long-lived storage of the information about the photon quantum state and its reconstruction.

We assume that the atoms are initially on the ground level $|1\rangle$. There are allowed electric dipole transitions $|1\rangle$ - $|3\rangle$ and $|1\rangle$ - $|2\rangle$, where $|2\rangle$ is a metastable level energetically close to $|1\rangle$. The frequency of the $|1\rangle$ - $|3\rangle$ transition is Doppler broadened. The atomic transitions from $|3\rangle$ to $|2\rangle$ levels will result only from independent spontaneous decays of single atoms. Thus, we can neglect by the effect of these spontaneous transitions on the photon absorption process in the chosen time scale. We also disregard the relaxation processes in the atomic system and interatomic collisions. Following [5], we also assume that it is possible to excite the transition $|3\rangle$-$|2\rangle$ by a laser pulse without perturbation of the atoms on the level $|1\rangle$.

Thus, we have the following initial state of the wave function ($t = -\infty$) in the interaction picture:

$$|\psi_f(t=-\infty)\rangle = \int dk f_k(-\infty) a_k^+ |0\rangle |A(t)\rangle. \tag{1}$$

Here $f_k(t)$ is the one-dimensional photon wave function ($\int dk |f_k(-\infty)|^2 = 1$) with maximum at $k = \vec{e}_z \omega_{ph}/c$ and the spectral width $\delta\omega_{ph}$, $|0\rangle$ - vacuum state of light, $a_k, a_k^+$ - the field operators, $|A(t)\rangle = \prod_{j=1}^{N} |\phi(t)\rangle_j$ is the atomic ground state, $|\phi(t)\rangle_j = \delta^{1/2}(z_j - z_j(t))|1\rangle_j$, $|1\rangle_j$ is the ground state of the j-th atom, the function $\delta^{1/2}(z_j - z_j(t))$ describes spatial movement of the j-th atom with coordinate $z_j(t) = z_j^o + v_z^j t$, $v^j$ - the velocity of j-th atom. In this one dimensional approach we describe our system behaviour using the following Hamiltonian:

$$H = H_a + H_f + V + V_1 + V_2; \tag{2}$$

$$H_a = \eta\omega_{31} \sum_{j=1}^{N} P_{33}^j + \eta\omega_{21} \sum_{j=1}^{N} P_{22}^j, \quad H_f = \int dk \, \eta\omega_k a_k^+ a_k, \tag{3}$$

$$V = \eta g \int dk \sum_{j=1}^{N} \{a_k P_{31}^j \exp(ikz_j) + h.c.\}, \tag{4}$$

$$V_m(t) = -\tfrac{1}{2}\eta\sum_{j=1}^{N}\Omega_m[(t-t_m-n_m z_j/c)/T_m]\{P_{32}^j \exp[-i(\omega_m t - k_m z_j)+i\varphi_m]+h.c.\}, \tag{5}$$

$\omega_{\nu\mu}$ - the central frequencies of the transitions $|\nu\rangle_j \to |\mu\rangle_j$; $P_{\nu\mu}^j$ are the j-atom operators coupling the states; m=1,2; $\Omega_m(t) = dE_m(t/T_m)/\eta$ - Rabi frequencies, $T_m$ - temporal duration of the m-th laser pulses; $d = d_{32} = d_{23}$ - dipole moment of the atomic transition $|2\rangle$ - $|3\rangle$; $E_m(t)$ is the amplitude of an electric field of the m-laser pulse; $g$ is the interaction constant of the photon with j-th atom. For the sake of simplicity we do not consider below the polarization aspects of the interaction of photons with the atoms.

$$|\psi(t)\rangle = |\psi_m(t)\rangle + |\psi_f(t)\rangle, \qquad (6)$$

$$|\psi_m(t)\rangle = |\psi_m^{(2)}(t)\rangle + |\psi_m^{(3)}(t)\rangle, \quad |\Psi_f(t)\rangle = \int dk f_k(t) a_k^+ |0\rangle |A(t)\rangle, \qquad (7)$$

$$|\psi_m^{(2)}(t)\rangle = \sum_{j=1}^N \xi_j(t) P_{21}^j |0\rangle |A(t)\rangle, \quad |\psi_m^{(3)}(t)\rangle = \sum_{j=1}^N b_j(t) P_{31}^j |0\rangle |A(t)\rangle \qquad (8)$$

Such structure of the wave function is determined by the initial conditions (1) and by the conservation of the total number excitations in total quantum system. After the substitution of (6)-(8) into the Schrödinger equation we obtain the following basic equations for the functions $f_k(t), b_j(t), \xi_j(t)$:

$$\tfrac{\partial}{\partial t} f_k(t) = -i\omega_k f_k(t) - ig \sum_{j=1}^N b_j(t) \exp\{-ikz_j(t)\} \qquad (9)$$

$$\begin{aligned}\tfrac{\partial}{\partial t} b_j(t) =\ & -i\omega_{31} b_j(t) - ig \int dk f_k(t) \exp\{ikz_j(t)\} \\
& + i\tfrac{1}{2}\xi_j(t)\{\Omega_1[(t-t_1-n_1 z_j(t)/c)/T_1]\exp[-i(\omega_1 t - k_1 z_j(t)) + i\varphi_1] + \\
& + \Omega_2[(t-t_2-n_2 z_j(t)/c)/T_2]\exp[-i(\omega_2 t - k_2 z_j(t)) + i\varphi_2]\}\end{aligned} \qquad (10)$$

$$\begin{aligned}\tfrac{\partial}{\partial t} \xi_j(t) =\ & -i\Delta\xi(t)_j + i\tfrac{1}{2} b_j(t)\{\Omega_1[(t-t_1-n_1 z_j(t)/c)/T_1]\exp[i(\omega_1 t - k_1 z_j(t)) - i\varphi_1] + \\
& + \Omega_2[(t-t_2-n_2 z_j(t)/c)/T_2]\exp[+i(\omega_2 t - k_2 z_j(t)) - i\varphi_2]\}\end{aligned} \qquad (11)$$

It is difficult to find a general solution of the equations (9)-(11), which include the EIT-effect as a special case. We consider a different physical situation when the photon wave packet and two laser pulses interact with medium at the different time moments. At this situation the quantum coherence at the interactions with the inhomogeneously broadened lines can be easy controlled experimentally in time, for we have to deal with quantum evolution of different type for different stages of the coherent interactions.

## 3. Mapping of the photon wave packet state

At this stage we study the photon absorption in the absence of the laser fields. Here the equations (9)-(10) are reduced to:

$$\tfrac{\partial}{\partial t} f_k = -i\omega_k f_k - ig \sum_{j=1}^N \beta_j \exp\{-ikz_j^o\}, \qquad (12)$$

$$\tfrac{\partial}{\partial t} \beta_j = -i\omega_{31}^{j(+)} \beta_j - ig \int dk f_k \exp\{ikz_j^o\}, \qquad (13)$$

where $\omega_{mn}^{j(+)} \equiv \omega_{mn}^j = \omega_{mn}(1 + v_z^j/c)$ and we used the substitution:

$$\beta_j = b_j(t)\exp\{-i(v_z^j/c)\omega_{31} t\}, \qquad (14)$$

Here we admit that $\exp\{i(k-\omega_{31}/c)v_z^j t\} \cong 1$, which takes place when $t < t_{max} = \omega_{31}/(\delta\omega_{ph}^2)$. The detailed analysis of conditions of validity of (12)-(13) will be presented in subsequent publications.

Multiplying (12) by $e^{ikz}$, integrating over $dk$ and using the notation $\int dk f_k(t) e^{ikz} = E(t,z)\exp\{-i\omega_{ph}(t-z/c)\}$, we obtain

$$(\tfrac{\partial}{c\partial t} + \tfrac{\partial}{\partial z}) E(t,z) = \exp\{\omega_{ph}(t-z/c)\} P(t,z), \qquad (15)$$

where $P(t,z) = -i(2\pi g/c)\sum_{j=1}^{N} \beta_j \delta(z - z_j^o)$.

We take into account all atoms localised in the photon propagation cross section with longitudinal coordinates close to "z" with their different velocities. In such way we define the summation over the arbitrary function $M(\omega_{31}^{j(+)}, t, z_j^o)$ with the $\delta(z - z_j^o)$-function using spectral inhomogeneous broadening $G((\omega - \omega_{31})/\Delta_n)$:

$$\sum_{j=1}^{N} M(\omega_{31}^{j(+)}, t, z_j^o) \delta(z - z_j^o) = n_o Q(z) \int_{-\infty}^{\infty} d\Delta G(\Delta/\Delta_n) M(\Delta + \omega_{31}, t, z). \qquad (16)$$

Here $\Delta = \omega_{31}^{j(+)} - \omega_{31}$, $n_o = N/L$ -atomic density, N – number of atoms; Q(z) reflects the spatial localization of the atoms in the gas tube: $Q(z) = 1$ {0<z<L} and Q(z)=0 {z<0;z>L}; $\omega_{31}^j - \omega_{31}$ - frequency detuning of atom-j. Using the formal solution of (13) we obtain the solution for the field E(t, z)

$$E(t,z) = \int_{-\infty}^{\infty} du \tilde{E}(u,z) \exp\{-iut\}. \qquad (17)$$

$$\tilde{E}(u,z) = \tilde{E}_o(u,0) \exp\{i\tfrac{u}{c} z - \alpha^{(+)}(u - \omega_{31} + \omega_{ph}) \int_{-\infty}^{Z} Q(z) dz\}, \qquad (18)$$

$$\tilde{E}_o(u,0) = \tfrac{1}{c} f_{(\omega_{ph} + u)/c}(-\infty). \qquad (19)$$

$$\alpha^{(+)}(u - \omega_{31} + \omega_{ph}) = \lim_{\varepsilon \to 0} \alpha_o \int_{-\infty}^{\infty} d\Delta' \frac{G(\Delta'/\Delta_n)}{\{\varepsilon + i[\Delta' - (u - \omega_{31} + \omega_{ph})]\}} \qquad (20)$$

where $\alpha_o = 2\pi n_o g^2/c$, and find the following solution for the wave function right after the emergence of the photon in the medium:

$$\left|\psi^i(t)\right\rangle\Big|_{t \gg \delta t_{ph}} = \left|\psi_m^i(t)\right\rangle + \left|\psi_f^i(t)\right\rangle,$$

$$\left|\psi_m^i(t)\right\rangle = \sum_{j=1}^{N} b_j(t) P_{31}^j |0\rangle A(t) , \left|\psi_f^i\right\rangle = \int dk f_k^{(1)}(t) a_k^+ |0\rangle A(t). \qquad (21)$$

It means that for the time $t \gg t_o = \max\{L/c, \delta t_{ph}\}$:

$$f_k^{(1)}(t) = \tfrac{1}{2\pi} e^{-i\omega_{ph} t} \int_{-\infty}^{\infty} dz E(t,z) \exp\{i(\omega_{ph}/c - k)z\}\Big|_{t/t_o \gg 1} \to f_k^{(1)}(\infty) \exp\{-i\omega_k t\}, \qquad (22)$$

$$f_k^{(1)}(\infty) = f_k(-\infty) \exp\{-\alpha^{(+)}(\omega_k - \omega_{ph}) L\}, \qquad (23)$$

$$b_j(t) = \beta(\omega_{31}^j; z_j^o) \exp\{-i(\omega_{31} t - \omega_{31}^{j(+)} z_j^o/c)\}, \qquad (24)$$

$$\beta(\omega_{31}^j; z_j^o) = -i 2\pi(g/c) f_{k=\omega_{31}^{j(+)}/c}(-\infty) \exp\{-\alpha^{(+)}(\omega_{31}^{j(+)} - \omega_{31}) z_j^o\}, \qquad (25)$$

Physically, the factor- $\exp\{-\alpha(\omega_{31}^{j(+)} - \omega_{31}) z_j^o + i\omega_{31}^{j(+)} z_j^o/c\}$ in (25) gives the probability amplitude for the photon to propagate as far as j-th atom, while the factor ( $2\pi g f_{k=\omega_{31}^{j(+)}/c}(-\infty)$ ) gives the probability amplitude of the photon absorption by the atom. Obviously that $\exp\{-\alpha^{(+)}(\omega_k - \omega_{31}) L\} \to 0$ is the sufficient condition for the pure mapping of the quantum state for a large class of quantum states of light with energy sufficiently small when Lambert-Bear absorption takes place.

Now we turn to times $t \gg t_{ph}$ when the field (17) $E(t,z) \to 0$ in the medium volume and consider the storage stage.

## 4. Storage of the mapped field state

We apply a laser pulse propagating in the same direction as the initial photon at the time delay $t = t_1 \ll T_2^{(3)}$ with the carrier frequency $\omega_1$ coinciding with $|3\rangle - |2\rangle$ transition and temporal profile $E_1(t') = E_{1,0}\text{Sech}\{t'/T_1\}$ of duration $T_1$. The basic equations (9)-(11) transform into the equations:

$$\tfrac{\partial}{\partial t}b_j(t) = -i\omega_{31}b_j(t) + i\tfrac{1}{2}\xi_j(t)\Omega_{1,0}\text{Sech}[(t-t_1-n_1z_j(t)/c)/T_1]\exp[-i(\omega_1 t - k_1 z_j(t)) + i\varphi_1], \quad (26)$$

$$\tfrac{\partial}{\partial t}\xi_j(t) = -i\Delta\xi(t)_j + i\tfrac{1}{2}b_j(t)\{\Omega_{1,0}\text{Sech}[(t-t_1-n_1z_j(t)/c)/T_1]\exp[i(\omega_1 t - k_1 z_j(t)) - i\varphi_1], \quad (27)$$

where $\Omega_{1,0} = dE_{1,0}/\eta$ with the initial conditions for (26), (27): $\xi_j(-\infty) = 0$ and the solution (24) for $b_j(t)$. Let us introduce the substitution:

$$\chi_j^m(t) = b_j(t)\exp\{i(\omega_{31}t - \phi_j^m/2)\};\ \varsigma_j^m(t) = \xi_j(t)\exp\{i(\Delta t + \phi_j^m/2)\}. \quad (28)$$

here m=1, denoting the first laser pulse. In the interaction picture we have the equations

$$\tfrac{\partial}{\partial t}\chi_j^m(t) = i\tfrac{1}{2}\varsigma_j^m(t)\Omega_{m,0}\text{Sech}[(t-t_j^m)/T_m^j]\exp\{-i\delta_j^m(t-t_j^m)\}, \quad (29)$$

$$\tfrac{\partial}{\partial t}\varsigma_j^m(t) = i\tfrac{1}{2}\chi_j^m(t)\Omega_{m,0}\text{Sech}[(t-t_j^m)/T_m^j]\exp\{i\delta_j^m(t-t_j^m)\}, \quad (30)$$

where

$T_m^j = T_m/(1-v_{n_m}^j/c)$, $n_m = k_m/|k_m|$, $\delta_j^m = \omega_m - \omega_{32} - \omega_m v_{n_m}^j/c$, $\omega_{32} = \omega_{31} - \Delta$,

$v_{n_m}^j = n_m v_j$, $|k_m| = \omega_m/c$, $\phi_j^m = k_m z_j^o + \varphi_m - \delta_j^m t_j^m$; $t_j^m = (t_m + n_m z_j^o/c)/(1-v_{n_m}^j/c)$. (31)

$\text{Lim}_{t\to-\infty}\chi_j^1(t) \to \chi_j^1(-\infty) = \beta(\omega_{31}^j; z_j)\exp\{-i\phi_j^1/2\}$, $\text{Lim}_{t\to-\infty}\varsigma_j(t) \to 0$

Following [12, 13] we find the exact solution of the equations (31), (32) for the time $(t-t_j^1)/T_1^j \gg 1$ in the laboratory frame of references :

$$b_j(t) = \beta(\omega_{31}^j; z_j)\,_2F_1(\theta_1^j/(2\pi), -\theta_1^j/(2\pi); \gamma_{j,1}^{(2)}; 1)\exp\{-i(\omega_{31}t - \omega_{31}^{j(+)}z_j^o/c)\}, \quad (32)$$

$\xi_j(t) = $

$= (2\pi g/c)f_{k=\omega_{31}^{j(+)}/c}(-\infty)\dfrac{\sin(\theta_1^j/2)}{ch(\pi\delta_1^j T_1^j/2)}\exp\{-\alpha(\omega_{31}^{j(+)} - \omega_{31})z_j^o - i\Delta t + i\mu_1(t_1; v_j; z_j^o)\}, \quad (33)$

where $_2F_1(a,b;c;z)$ -the hypergeometric function; $\theta_m^j = \pi\Omega_{m,0}T_m^j$ - the pulse area $\gamma_m^{(2)} = 1/2 + i\delta_m^j T_m^j/2$, $\gamma_m^{(1)} = (\gamma_m^{(2)})^*$.

$\mu_1(t_1; v_j; z_j^0) = \omega_1 t_1 - \varphi_1 - \omega_{32}(1+v_{n_1}^j/c)t_1 + \delta\mu_1(z_j^0)$,

$\delta\mu_1(z_j^0) = \omega_{31}(1+v_z^j/c)z_j^0/c - \omega_{32}(1+v_{n_1}^j/c)n_1 z_j^0/c$. (34)

Note the function $_2F_1(\theta_1^j/(2\pi), -\theta_1^j/(2\pi); \gamma_{j,1}^{(2)}; 1)$ is spatially and temporally independent, thus the laser pulse only decreases the amplitude $b_j(t)$ independently of the spatial position of j-th atom. The amplitudes $b_j(t)$ will be close to zero for the values $\theta_1^j \approx \pi$ and $\gamma_{j,1}^{(2)} \approx 1/2$ which takes place only for atoms with spectral detunings $\delta_1^j < (2/\pi T_1)$. It means the possibility of spectral control to realise the independent quantum storage for a narrow spectral group of atoms within inhomogeneously broadened transition 2-3 ($(2/\pi T_1) \ll \Delta_n$).

A new spatial and temporal behaviour takes place for the coefficients $\xi_j(t)$ which are weakly affected by slow atomic movement since the main information about the quantum state of the photon is kept in the spectral properties of the atomic excitation, which is a distinctive feature of the proposed technique as compared to the EIT technique. The advantage may be of crucial importance for the QM realization using short pulses of light.

## 5. Reconstraction of the single photon state

In this section we consider the spontaneous emission of the atomic gas after its excitation by the second restoring laser pulse. The pulse propagates in the opposite direction with respect to the initial photon. In this case, any j-th atom having the velocity $v_z$ changes the sign of its frequency detuning on the opposite, which makes possible the rephasing of the macroscopic atomic coherence induced by the initial photon. The wave function right after the second laser pulse can be found solving the equations similar to (26), (27) with the index m equal to 2. The main information sufficient for study of the quantum state reconstruction can be obtained using the solution for the atomic coefficients $b_j(t)$. This solution has the form:

$$b_j(t) = (b_j^{(1)} + b_j^{(2)})e^{-i\omega_{31}t}, \qquad (35)$$

where

$$b_j^{(1)} = {}_2F_1(\theta_1^j/(2\pi), -\theta_1^j/(2\pi); \gamma_{j,1}^{(2)};1) \, {}_2F_1(\theta_2^j/(2\pi), -\theta_2^j/(2\pi); \gamma_{j,2}^{(2)};1) \, \beta(\omega_{31}^j;z_j)$$
$$\exp\{i\omega_{31}^{j(+)}z_j^o/c\} \qquad (36)$$

$$b_j^{(2)} = (2\pi i g/c) \frac{\sin\theta_1^j/2}{\operatorname{ch}(\pi\delta_1^j T_1^j/2)} \frac{\sin\theta_2^j/2}{\operatorname{ch}(\pi\delta_2^j T_2^j/2)} f_{k=\omega_{31}^{j(+)}/c}(-\infty) \exp\{-\alpha(\omega_{31}^{j(+)}-\omega_{ph})z_j\} \exp\{i\Psi(t_1,t_2;z_j^o)\}, \qquad (37)$$

where

$$\Psi(t_1,t_2;z_j^o) = \omega_{32}(t_2 - t_1) - (2\omega_{32} - \omega_{31})z_j^o/c + (v_z^j/c)[\omega_{31}z_j^o/c - \omega_{32}(t_1+t_2)] + \varphi_{se}. \qquad (38)$$

and $\varphi_{se} = \omega_1 t_1 - \omega_2 t_2 + \varphi_{21}$.

The first term $b_j^{(1)}(t)$ determines the behavior of the atomic system which took place right after the absorption of the photon, but its behavior is suppressed by the action of the two laser pulses. Obviously, the term (36) will not significantly contribute to the coherent interaction with the quantum field which must be neglected. At the selected parameters of the light fields the atomic system will coherently irradiate the photon in the backward direction with respect to the initial photon $k_{se}\uparrow\downarrow k_{ph}$ [11]. Therefore, we use new notation for the field modes $\vec{k}' = -k\vec{e}_z$, $\omega_{-k} = \omega_k$ and the equations (9)-(11) take the form:

$$\tfrac{\partial}{\partial t}f_{-k}(t) = -i\omega_k f_{-k}(t) - ig\sum_{j=1}^{N} b_j(t)\exp\{ikz_j(t)\}, \qquad (39)$$

$$\tfrac{\partial}{\partial t}b_j(t) = -i\omega_{31}b(t)_j - ig\int dk f_{-k}(t)\exp\{-ikz_j(t)\}, \qquad (40)$$

Similarly to (12)-(13), using the substitution $\eta_j = b_j(t)\exp\{i(v_z^j/c)\omega_{31}t\}$ we have

$$\tfrac{\partial}{\partial t}f_{-k} = -i\omega_k f_{-k} - ig\sum_{j=1}^{N}\eta_j e^{ikz_j^o}, \qquad (41)$$

$$\tfrac{\partial}{\partial t}\eta_j = -i\omega_{31}^{j(-)}\eta_j - ig\int dk f_{-k}e^{-ikz_j^o} \qquad (42)$$

with the initial quantum states

$$f_k(t = t_1 = T + \tau + \delta t_2) = 0 \qquad (43)$$

$$\eta_j(t = t_1 = T + \tau + \delta t_2) = \eta_j^{(2)}(t_1), \qquad (44)$$

here

$$\eta_j^{(1),(2)}(t_1) = b_j^{(1),(2)}\exp\{-i(1 - v_z^j/c)\omega_{31}t_1\}. \qquad (45)$$

Following the mathematical procedures made in the mapping section we introduce the notation $\int dk f_{-k}e^{-ikz} = F(t,z)\exp\{-i\omega_{ph}'(t+z/c)\}$, where $\omega_{ph}'$, $F(t,z)$, are unknown carrying frequency and envelope of the irradiated field. From (41), (42) we obtain the following equation:

$$(\tfrac{\partial}{c\partial t} - \tfrac{\partial}{\partial z})F(t,z) = \exp\{i\omega_{ph}'(z/c + t)\}(P_2(t,z) + P_3(t,z)), \qquad (46)$$

here

$$P_2(t,z) = -i(2\pi g/c)\sum_{j=1}^{N}\eta_j^{(2)}(t_1)\exp\{-i\omega_{31}^{j(-)}(t-t_1)\}\delta(z-z_j^o), \qquad (47)$$

$$P_3(t,z) = -(2\pi g^2/c)\int_{t_1}^{t}dt'\sum_{j=1}^{N}F(t',z_j^o)\exp\{-i[\omega_{31}^{j(-)}(t-t') + \omega_{ph}'(z_j^o/c + t')]\}\delta(z-z_j^o), \qquad (48)$$

here $\omega_{mn}^{j(-)} \equiv \omega_{mn}^{j} = \omega_{mn}(1 - v_z^j/c)$.

Below we dwell upon on the case of equal pulse areas $\theta_1 = \theta_2 = \theta$ and laser pulse durations $T_1 = T_2 = T$, and consider the case $\omega_1 = \omega_2 = \omega_{32}$. After summation over the atoms we have:

$$(\tfrac{\partial}{c\partial t} - \tfrac{\partial}{\partial z})F(t,z) = Y(t,z) \qquad (49)$$

$$Y(t,z) = i(2\pi n_o g/c)Q(z)\exp[i(\omega_{ph}' - \omega_{31})t - i(2\omega_{32} - \omega_{31} - \omega_{ph}')z/c]\exp\{i\varphi_{2,1}\}$$

$$\int_{-\infty}^{+\infty} d\Delta G(\Delta/\Delta_n)\beta(\Delta + \omega_{31};z)\frac{\sin^2(\theta/2)}{\cosh^2\{\pi(\omega_{32}/\omega_{31})\Delta T/2\}}\exp\{i\Delta[t - (\omega_{32}/\omega_{31})(t_1 + t_2) + z_j^o/c]\} -$$

$$-\alpha_o Q(z)\int_{-\infty}^{t}dt'F(t',z)\exp\{-i(\omega_{31} - \omega_{ph}')(t-t')\}\int_{-\infty}^{\infty}d\Delta G(\Delta/\Delta_n)\exp\{i\Delta(t-t')\} \qquad (50)$$

which solution has the form $F(t,z) = \int_{-\infty}^{\infty}du\tilde{F}_o(u,z)\exp\{-iu(t+z/c)\}$:

$$\tilde{F}_o(u,z) =$$

$$= \frac{\sin^2(\theta/2)}{\cosh^2\{\pi(\omega_{32}/\omega_{31})(\omega_{31} - \omega_{ph}' - u)T/2\}}\frac{1}{(1 - i\omega_{21}/[\pi c\alpha_o G(\omega_{31} - \omega_{ph}' - u)])}\tfrac{1}{c}f_{(2\omega_{31} - \omega_{ph}' - u)/c}(-\infty)$$

$$\{\exp[i2\omega_{21}z/c - \alpha^{(+)}(\omega_{31} - \omega_{ph}' - u)z] - \exp[i2\omega_{21}L/c - \alpha^{(+)}(\omega_{31} - \omega_{ph}' - u)L]\}$$

$$\exp[-i(\omega_{31} - \omega_{ph}' - u)(\omega_{32}/\omega_{31})(t_1 + t_2)]\exp\{i[\varphi_{2,1} + \overline{\Delta}(t_1 + t_2)]\} \qquad (51)$$

The field wave function

$$f_{\omega_{-k}/c}^{(2)}(t \to \infty) = c\tilde{F}_o(\omega_{-k} - \omega_{ph}', 0) \qquad (52)$$

with the general state field + atoms in the form (21), where

$$|\psi_f^i\rangle = \int_0^{\infty}dk\{f_{\omega_k/c}^{(1)}(t \to \infty)a_k^+ + f_{\omega_{-k}/c}^{(2)}(t \to \infty)a_{-k}^+\}|0\rangle|A(t)\rangle. \qquad (53)$$

We confine ourselves to the analysis of the irradiated field variables. Using the obtained solution (51)-(53) for the wave function we can analyze it in the particular cases.

## 6. Analysis of the reconstructed field and conclusion

As is seen from (51) the perfect reconstruction of the photon state requires that

$$\omega_{21}/[\pi c \alpha_o G(\omega_{31}^{j(+)} - \omega_{31})] \ll 1 \qquad (54)$$

Under this favorite conditions and optically dense medium $\exp\{-\alpha^{(+)}(\omega_k - \omega_{31})L\} \to 0$ for all Fourier components of the photon field the state of the restored radiation becomes connected with the initial state of the photon in the following way:

$$f^{(2)}_{\omega_{-k}/c}(t \to \infty) = \exp\{-i(\omega_{-k}t - \Phi_{-k})\} f_{(2\omega_{31}-\omega_k)/c}(-\infty) , \qquad (55)$$

where $\Phi_{-k} = (\omega_{-k} - \omega_{31})(\omega_{32}/\omega_{31})(t_1 + t_2) + \varphi_{2,1})$, $(\omega_{-k} = \omega_k)$. That means the complete reconstruction of the absorbed photon.

It is interesting to analyze the case of the maximum spectrally broadened photon wave function which can be reconstructed with high efficiency. At this case we have to consider the medium with arbitrary optical density, length spectral broadening. The characteristic dependences of the main solution (51) for some special cases are shown in Figs. 2-5. In numerical calculations we took into account the Lorentz spectral profiles of the inhomogeneous broadening $G(\Delta/\Delta_n) = \frac{\Delta_n}{\pi(\Delta_n^2 + \Delta^2)}$ and of the photon wave function $|f_{(\omega_{31}+\Delta)/c}(-\infty)|^2 = \frac{\delta\omega_{ph}}{\pi(\delta\omega_{ph}^2 + \Delta^2)}$. The numerical results in Fig.2 demonstrate that the reconstruction of spectral components of the initial photon wave function at the fixed parameters $\delta\omega_{ph}= 2\cdot 10^8$ sec$^{-1}$, $\omega_{21}=10^{10}$ sec$^{-1}$, $\alpha=1$ cm$^{-1}$ and $\Delta_n = 10^9$ sec$^{-1}$. At this parameters the complete reconstruction takes place for L > 3.

Nonsymmetrical spectral reconstruction is shown in Fig.3. This spectral deformation of the initial photon spectrum is determined by the spectral dispersion effect within the inhomogeneous broadening. This effect increases with sufficiently large spectral width of the photon wave function and small optical depth $\alpha L \leq 1$. In this case the total probability of the photon reconstruction is about 0.23. The behavior of the total probability of the photon reconstruction is studied in Figs. 4,5. The probability is high for the frequencies $\omega_{21}=\cdot 10^{10}$ at $\Delta_n = 10^9$ sec$^{-1}$, $\delta\omega_{ph}= 2\cdot 10^8$ and $\alpha L \geq 3$. It is noteworthy that the splitting between the first and the second atomic levels may exceed the spectral widths of the photon wave function and the inhomogeneous broadening. It makes possible the spectral selection of the weak quantum field from the laser pulses.

In conclusion, we note the obtained basic relationships describe the possibilities of the model three-level atomic system. It is quite possible that practical realization of this photon echo technique will require a switch-over to more sophisticated atomic systems with four and more working levels.

This work was supported by the Royal Swedish Academy of Sciences, the Russian Foundation for Basic Research grants No.: 01-01-00387, 01-03-32730, 02-03-06708 mac.

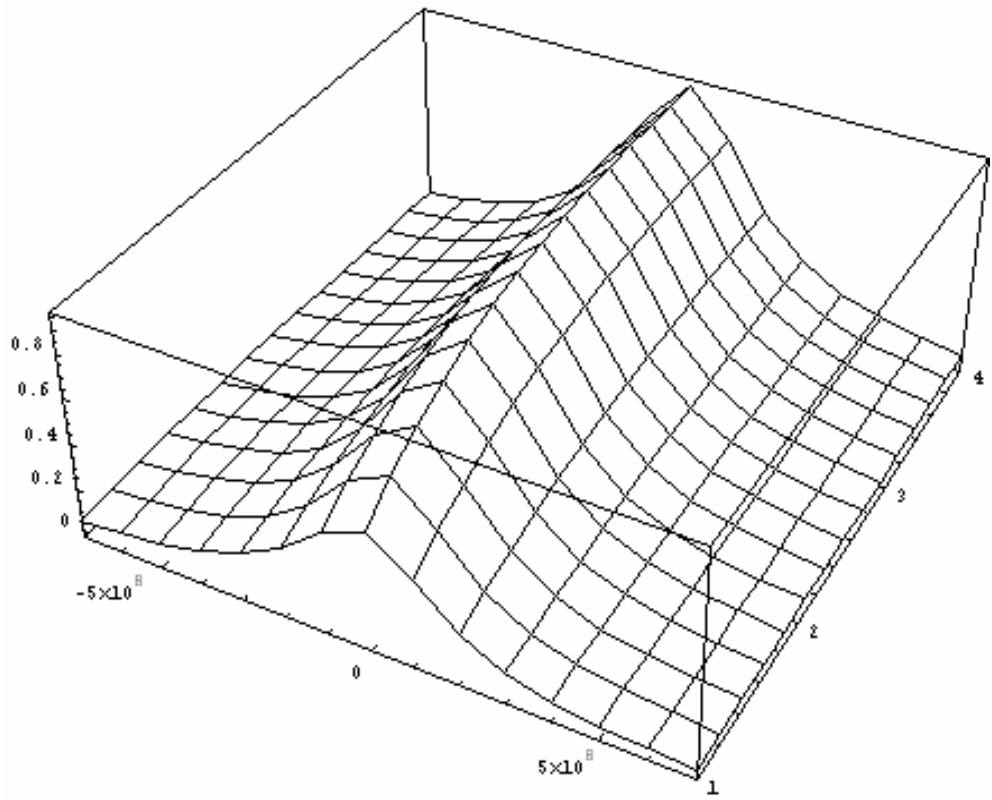

Fig 2.
S.A.Moiseev, et.al. The possibilities of the quantum memory realization for short pulses of light in the photon echo technique

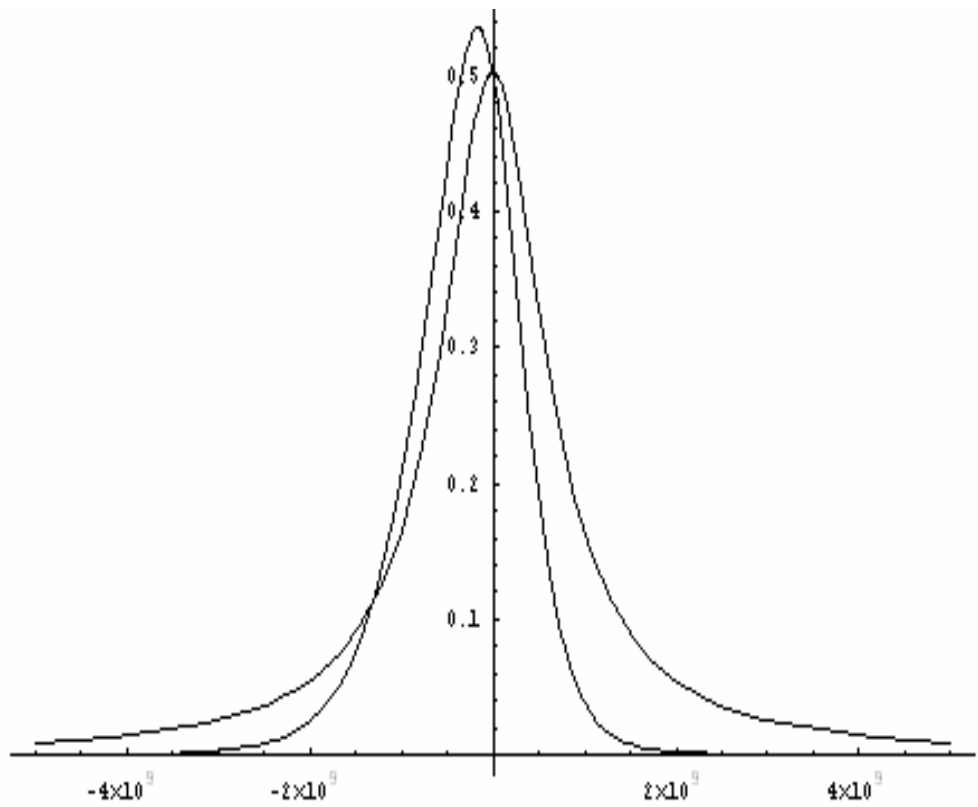

Fig.3.
S.A.Moiseev, et.al. The possibilities of the quantum memory realization for short pulses of light in the photon echo technique

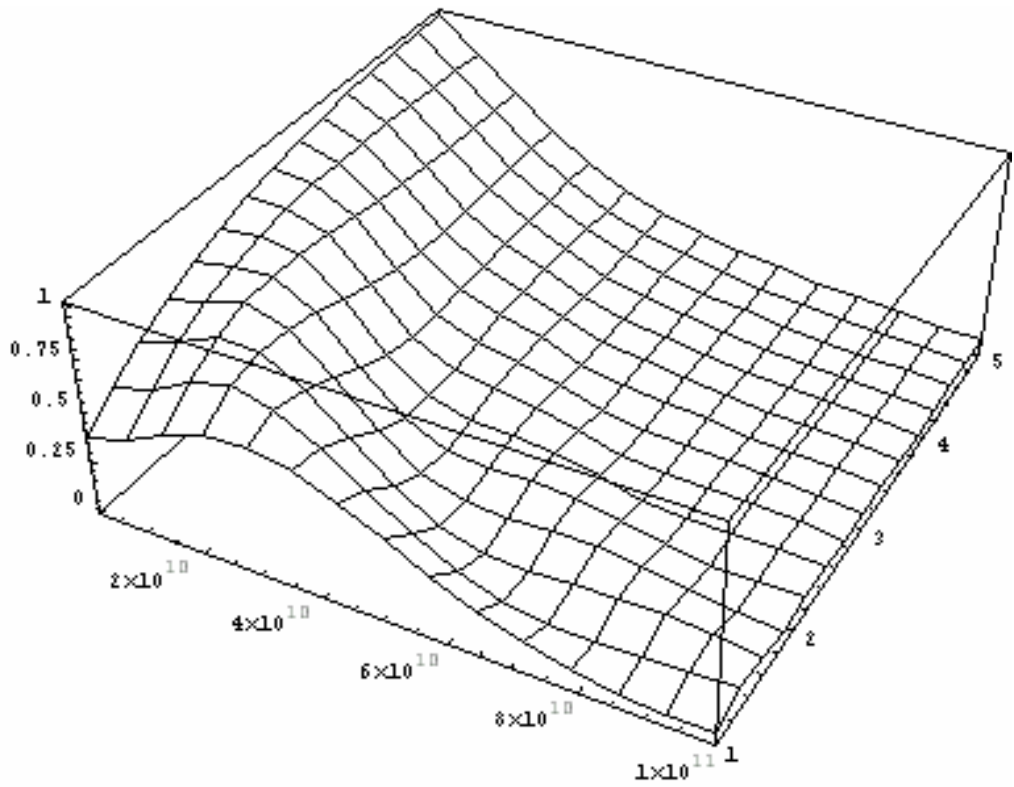

Fig. 4
S.A.Moiseev, et.al. The possibilities of the quantum memory realization for short pulses of light in the photon echo technique

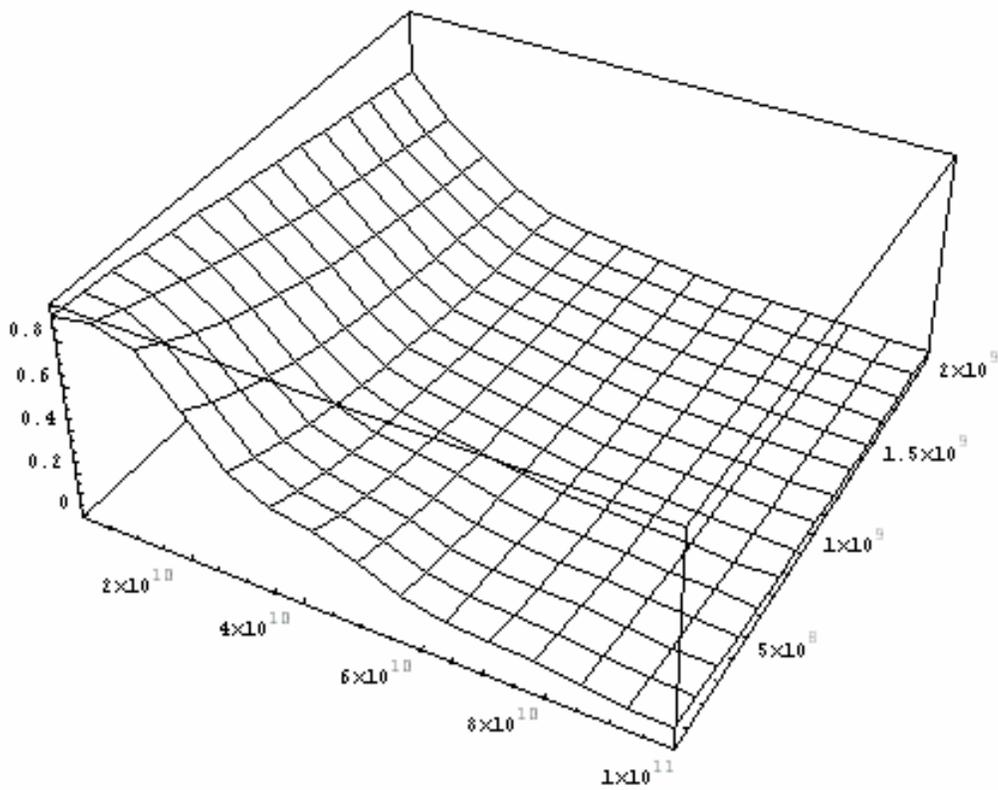

Fig. 5
S.A.Moiseev, et.al. The possibilities of the quantum memory realization for short pulses of light in the photon echo technique

Fig.1. The temporal diagram of the nonstationary interactions and the scheme of energetic levels. The first line from the left represents the photon entering the medium in the resonance with atomic transition $|1\rangle$-$|3\rangle$, after time delay $\tau$ the laser pulse transfers the atomic excitation from level $|3\rangle$ onto the level $|2\rangle$, and after delay T – the second laser pulse returns the excitation from $|2\rangle$ onto atomic level $|3\rangle$ with an additional atomic phase. The reconstructed photon is irradiated in the echo signal with time delay $\tau$ after the second laser pulse. The first laser pulse propagates in the photon wave packet direction ($k_1 \uparrow\uparrow k_{ph}$), while the second laser pulse propagates in the opposite direction ($k_2 \uparrow\downarrow k_{ph}$). Photon echo signal is emitted in the direction $k_e \uparrow\uparrow k_2$).

Fig 2. The spectral function $\left|f^{(2)}_{(\omega_{31}-\Delta)/c}(t\to\infty)\right|^2 / \left|f_{\omega_{31}/c}(-\infty)\right|^2$ of the reconstructed photon is shown as a function of the spectral detuning $\Delta$ and optical depth of the medium L ($\Delta$: -$5 \cdot 10^8 \text{sec}^{-1} \div 5 \cdot 10^8 \text{sec}^{-1}$, L: $1 \div 4$ cm) at the following fixed parameters $\Delta_n = 10^9 \text{sec}^{-1}$, $\delta\omega_{ph}= 2 \cdot 10^8$ sec$^{-1}$, $\omega_{21}=10^{10}$ sec$^{-1}$, $\alpha=1$ cm$^{-1}$ and inhomogeneous broadening $G(\Delta/\Delta_n) = \frac{\Delta_n}{\pi(\Delta_n^2+\Delta^2)}$. The function is normalized with respect to the center of the spectrum of the initial photon, which is taken in the form $\left|f_{(\omega_{31}+\Delta)/c}(-\infty)\right|^2 = \frac{\delta\omega_{ph}}{\pi(\delta\omega_{ph}^2+\Delta^2)}$. We see the complete photon reconstruction for L > 3 ($\alpha=1$ cm$^{-1}$).

Fig.3. The nonsymmetrical spectral reconstruction of the spectrum for the small optical depth L= 1, $\alpha=1$ cm$^{-1}$, $\Delta_n = 10^9 \text{sec}^{-1}$, $\delta\omega_{ph}= 7 \cdot 10^8$ sec$^{-1}$, $\omega_{21}=10^{10}$ sec$^{-1}$, and inhomogeneous broadening $G(\Delta/\Delta_n) = \frac{\Delta_n}{\pi(\Delta_n^2+\Delta^2)}$ and the initial spectrum $\left|f_{(\omega_{31}+\Delta)/c}(-\infty)\right|^2 = \frac{\delta\omega_{ph}}{\pi(\delta\omega_{ph}^2+\Delta^2)}$, normalized as in Fig.3.

Fig. 4 The total probability of the photon reconstruction as a function of medium length L and frequency splitting $\omega_{21}$ at the following fixed parameters $\Delta_n = 10^9 \text{sec}^{-1}$, $\delta\omega_{ph}= 2 \cdot 10^8$ sec$^{-1}$, $\alpha=1$ cm$^{-1}$.

Fig. 5 The total probability of the photon reconstruction as a function of the frequency splitting $\omega_{21}$ and spectral width $\delta\omega_{ph}$ of the initial photon wave function at the following fixed parameters $\Delta_n = 10^9 \text{sec}^{-1}$, $\alpha=1$ cm$^{-1}$. L=1 cm, ($\omega_{21}$ : $10^{10}$ sec$^{-1} \div 10^{11}$ sec$^{-1}$; $\delta\omega_{ph}$: $10^8$ sec$^{-1} \div 2 \cdot 10^9$ sec$^{-1}$.

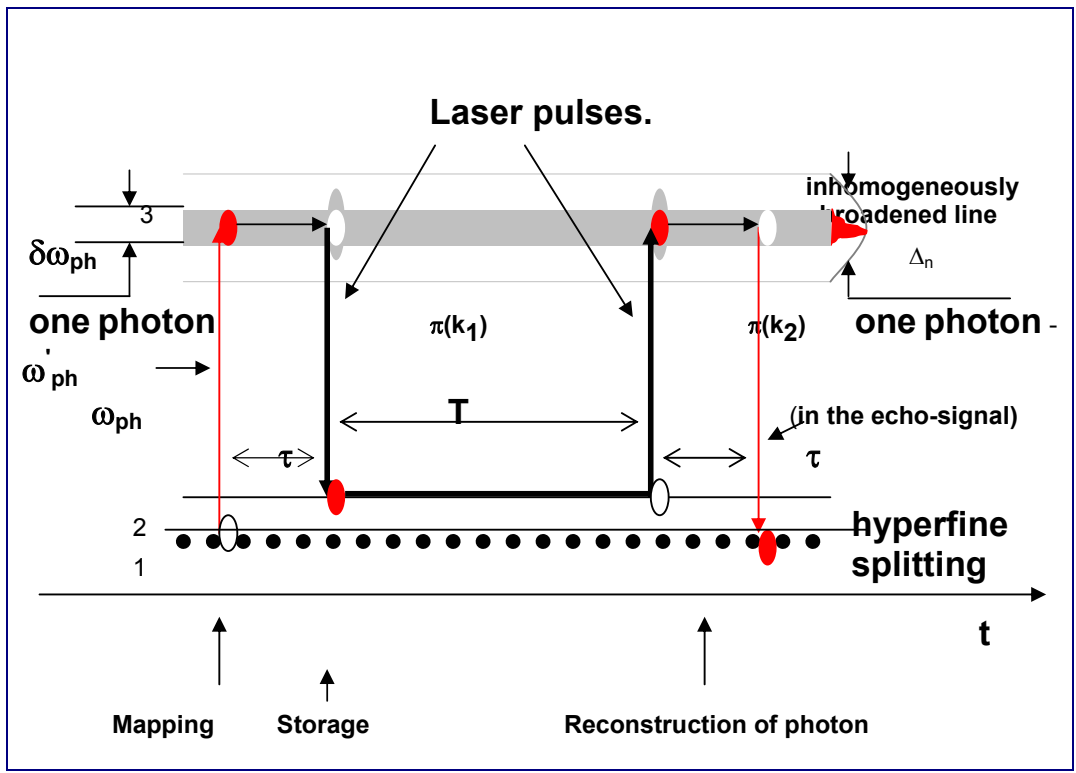

Fig. 1
S.A.Moiseev, et.al. The possibilities of the quantum memory realization for short pulses of light in the photon echo technique